\documentclass[prc,superscriptaddress,unsortedaddress,showpacs]{revtex4}	

\usepackage{graphicx}
\usepackage{amsmath}
\usepackage{amssymb}
\usepackage{times}
\usepackage{comment}
\usepackage{bm}
\usepackage{braket}
\usepackage{color}
\usepackage{ulem}

\def\fun#1#2{\lower3.6pt\vbox{\baselineskip0pt\lineskip.9pt
\ialign{$\mathsurround=0pt#1\hfil##\hfil$\crcr#2\crcr\sim\crcr}}}

\usepackage{soul}

\begin{document}

\title{ Manifestation of $\alpha$-clustering in ${}^{10}$Be via
$\alpha$-knockout reaction }

\author{Mengjiao Lyu} \email[]{mengjiao@rcnp.osaka-u.ac.jp}
\affiliation{Research Center for Nuclear Physics (RCNP), Osaka University,
Ibaraki 567-0047, Japan}

\author{Kazuki Yoshida} \email[]{yoshidak@rcnp.osaka-u.ac.jp}
\affiliation{Research Center for Nuclear Physics (RCNP), Osaka University,
Ibaraki 567-0047, Japan}

\author{Yoshiko Kanada-En'yo} \email[]{yenyo@ruby.scphys.kyoto-u.ac.jp}
\affiliation{Department of Physics, Kyoto University, Kyoto 606-8502, Japan}

\author{Kazuyuki Ogata} \email[]{kazuyuki@rcnp.osaka-u.ac.jp}
\affiliation{Research Center for Nuclear Physics (RCNP), Osaka University,
Ibaraki 567-0047, Japan}

\date{\today}

\begin{abstract}

\begin{description}
\item[Background] Proton-induced $\alpha$-knockout reactions may allow direct
experimental observation of $\alpha$-clustering in nuclei. This is obtained by
relating the theoretical descriptions of clustering states with experimental
reaction observables. It is desired to introduce microscopic structure models
into the theoretical frameworks for $\alpha$-knockout reactions.

\item[Purpose] Our goal is to probe the $\alpha$-clustering in ${}^{10}$Be
nucleus by proton-induced $\alpha$-knockout reaction observables.

\item[Method] We adopt an extended version of the
Tohsaki-Horiuchi-Schuck-R{\"o}pke (THSR) wave function of ${}^{10}$Be and
integrate it with the distorted wave impulse approximation (DWIA) framework for
the calculation of $(p,p\alpha)$ knockout reactions.

\item[Results] We make the first  calculation for the
${}^{10}$Be(p,p$\alpha$)${}^{6}$He reaction at 250 MeV implementing a
microscopic $\alpha$-cluster wave function and predict the triple differential
cross sections (TDX). Furthermore, by constructing artificial states of the
target nucleus ${}^{10}$Be with compact or dilute spatial distributions, the TDX
is found to be highly sensitive to the extent of clustering in the target
nuclei.

\item[Conclusions] These results provide reliable manifestation of the
$\alpha$-clustering in ${}^{10}$Be.

\end{description}
\end{abstract}


\maketitle

\section{Introduction} 

Clustering is one of the fundamental degrees of freedom in nuclei
\cite{horiuchi12} that originates from the delicate balances between Pauli
blocking effects and nucleon-nucleon interactions in nuclear many-body dynamics
\cite{freer17}. Hence, a microscopic description, which takes into account both
the nucleon degrees of freedom in interactions and the total antisymmetrization,
is essential for the study of nuclear clustering states. In recent years, the
microscopic theories have been well established for clustering states ranging
from the molecular-like states in ${}^{9,10}$Be~\cite{Oka77, Sey81, Des89,
Oer96,Ara96, Dot97, Kan99,Oga00, itagaki00, Des02, Ito04,Oer06, kobayashi12,
Ito14,lyu15, lyu16} to the gas-like Hoyle state (${0}^{+}_2$) in ${}^{12}$C
\cite{tohsaki01,Fun15, Fun15a,Fun16, zhou16}. In these studies, physical
observables such as energies and radii are calculated for the clustering states
and the corresponding experimental values are well reproduced. This indicates
the validity of the $\alpha$-clustering picture for these states. However,
physical observables that are directly related to the cluster degree of freedom
will be necessary for an evident manifestation of clustering in nuclei.

It is expected that nuclear reactions with adding or removing
$\alpha$-cluster(s) provide direct probe of $\alpha$-clustering in
nuclei~\cite{fukui16,yoshida16,yoshida17}. In recent work, microscopic cluster
model based on the generator coordinate method has been introduced into the
theoretical framework of $\alpha$-transfer reactions and it significantly
improves the prediction of transfer cross sections~\cite{fukui16}. The
$\alpha$-knockout reaction, which is an alternative approach, has been adopted
for the investigation of $\alpha$-clustering in stable nuclei for
decades~\cite{Roos77,Nadasen80,Carey84,Wang85,
Nadasen89,Mabiala09,yoshida16,yoshida17,wakasa17}. In these studies, the
distorted wave impulse approximation (DWIA) has been adopted by employing
phenomenological $\alpha$-cluster wave functions. Using a hydrogen target, the
($p,p\alpha$) reaction can be applied to studies on $\alpha$-clustering in
unstable nuclei, which is a hot subject in nuclear
physics~\cite{Oer06,EH01,Kan12,Hor12}. In a recent theoretical
study~\cite{yoshida16}, the so-called factorization approximation which has
frequently been made in DWIA framework is validated. Furthermore, corresponding
DWIA calculations with phenomenological $\alpha$-cluster wave functions have
shown the peripheral property of the $\alpha$-knockout reaction, which lays the
foundations for directly probing $\alpha$-clustering in the surface region of
nuclei~\cite{yoshida16,yoshida17}.  Hence, it is appealing and promising to
integrate the microscopic clustering models into this DWIA framework and make
microscopic predictions for the $\alpha$-knockout reaction observables.

In the present work, we study the $\alpha$-knockout reaction
${}^{10}$Be($p$,$p\alpha$)${}^{6}$He at 250 MeV. The microscopic cluster models
for the ${}^{10}$Be nucleus are already well established and its ground state is
predicted to be   molecular-like. Along this line, measurement of proton-induced
$\alpha$-knockout reactions for Be isotopes in inverse kinematics is planned at
RIBF~\cite{yan17}. Thus, the ${}^{10}$Be($p$,$p\alpha$)${}^{6}$He knockout
reaction at 250~MeV will be an ideal choice for the manifestation of
$\alpha$-clustering. In our calculations, we adopt the extended version of the
Tohsaki-Horiuchi-Schuck-R{\"o}pke (THSR) wave function~\cite{tohsaki01} for the
description of target nucleus ${}^{10}$Be and the residual nucleus ${}^{6}$He
\cite{lyu16,zhao17}, and then integrate it into the DWIA framework to predict
the ${}^{10}$Be($p$,$p\alpha$)${}^{6}$He reaction observables. Furthermore,
benefiting from the flexible model space in the THSR wave function, we can
smoothly evolve the physical ground state of ${}^{10}$Be into artificial states
of cluster gas-like limit with large spatial spread or  compact
SU(3)-shell-model limit. This is obtained by controlling one parameter for the
$\alpha$-cluster distribution size (the $\alpha$-cluster motion) in the THSR
wave function. By investigating the dependence of reaction observables on the
$\alpha$-cluster motion, it is presented that the $\alpha$-knockout reaction is
a sensitive probe to distinguish the shell-model limit, the molecular-like, or
the gas-like cluster state, and can thus provide direct manifestation of nuclear
structures in $^{10}$Be.

This article is organized as follows. In Section II, we introduce the
theoretical framework for the $\alpha$-knockout reaction: the DWIA for
calculating transition amplitudes and triple differential cross sections (TDX),
and the THSR wave function for the microscopic description of target and
residual nuclei. In Section III, we show the numerical inputs and discuss the
results of both structures and reaction observables. Last Section IV contains
the conclusion.

\section{Formalism}
\label{secformalism}

We integrate the DWIA framework and the THSR wave function to formulate a
microscopic description of the ${}^{10}$Be($p$,$p\alpha$)${}^{6}$He knockout
reaction. 

\subsection{DWIA framework for the ${}^{10}$Be($p$,$p\alpha$)${}^{6}$He reaction}
\label{subsecdwia}

\begin{figure}[htbp]
\begin{center}
 \includegraphics[width=0.35\textwidth]{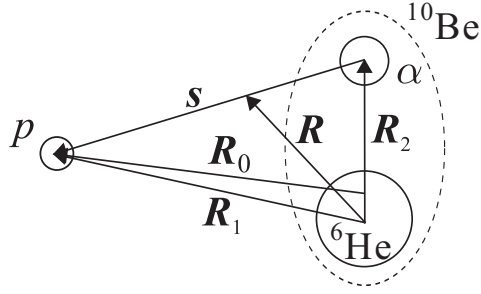}
 \caption{ Coordinates of the ${}^{10}$Be($p$,$p\alpha$)${}^{6}$He reaction. }
 \label{figcoord}
\end{center}
\end{figure}
We adopt the normal kinematics for the ${}^{10}$Be($p$,$p\alpha$)${}^{6}$He
reaction in the DWIA framework. The transition amplitude is given by
\begin{align}
&T_{{\bm K}_0{\bm K}_1{\bm K}_2}
= \nonumber \\
&\left<
\chi_{1,{\bm K}_1}^{(-)}({\bm R}_1)\chi_{2,{\bm K}_2}^{(-)} ({\bm R}_2)
\left| t_{p\alpha}({\bm s}) \right|
\chi_{0,{\bm K}_0}^{(+)}({\bm R}_0) \varphi_{\alpha}({\bm R}_2)
 \right>.
 \label{kotmtx}
\end{align}
Here, the incident proton $p$, the outgoing $p$, and the outgoing $\alpha$ are
labeled by 0, 1, and 2, respectively, and the distorted waves $\chi$ are
specified with these numbers in subscripts. The coordinates are given in
Fig.~\ref{figcoord}. The momentum and { its} solid angle of each particle are
denoted by ${\bm K}_i$ and $\Omega_i$ ($i=0,1,2$), respectively. Quantities with
superscript L are evaluated in the laboratory frame, and others are evaluated in
the center-of-mass frame. The superscripts $(+)$ and $(-)$ on $\chi$ indicate
that the outgoing- and incoming-wave boundary conditions are adopted,
respectively. The optical potential for each system is given by folding density
distributions of ${}^{10}$Be and ${}^{6}$He which are predicted by the THSR wave
function (see Sec.~\ref{thsr}) with an effective nucleon-nucleon (NN)
interaction. The three distorted wave functions are then obtained by solving the
corresponding Schr{\"o}dinger equations. $\varphi_{\alpha}$ is the
$\alpha$-cluster wave function inside ${}^{10}$Be nucleus with total angular
momentum $j=0$ and parity $\pi=+$. This $\alpha$-cluster wave function is
extracted from the THSR wave function of ${}^{10}$Be by approximating the
reduced width amplitude (RWA). Hence, all the wave functions used in the
calculation of transition amplitude are microscopically obtained. The transition
interaction $t_{p\alpha}$ between $p$ and $\alpha$ is obtained by a folding
model calculation as in Ref.~\cite{yoshida16}.

As illustrated in { Ref.~\cite{yoshida16}}, with the asymptotic momentum
approximation, the reduced transition amplitude $\bar{T}_{{\bm K}_0{\bm K}_1{\bm
K}_2}$ can be expressed as
\begin{align}
    \bar{T}_{{\bm K}_0{\bm K}_1{\bm K}_2}
=&
    {} \int d{\bm R}\, 
        F_{{\bm K}_0 {\bm K}_1 {\bm K}_2}({\bm R})\,\varphi_{\alpha}({\bm R}),
\label{localtmt}
\end{align}
where $F_{{\bm K}_0 {\bm K}_1 {\bm K}_2}({\bm R})$ is defined by
\begin{align}
    F_{{\bm K}_0 {\bm K}_1 {\bm K}_2}({\bm R})
&\equiv
    \chi_{1,{\bm K}_1}^{*(-)}({\bm R})\, 
    \chi_{2,{\bm K}_2}^{*(-)}({\bm R}) \nonumber \\
&\,\,\,\,\, \times 
    \chi_{0,{\bm K}_0}^{(+)}({\bm R})\,
    e^{-i{\bm K}_0 ({\bm R})\cdot {\bm R}(4/10)}.
\end{align}
With { $\bar{T}_{{\bm K}_0{\bm K}_1{\bm K}_2}$ in Eq.~(\ref{localtmt})}, the
triple differential cross section (TDX) of the
${}^{10}$Be($p,p\alpha$)${}^{6}$He reaction is given by
\begin{align}\label{eq:tdx}
    \frac{d^3 \sigma}{dE_1^{\rm L} d\Omega_1^{\rm L} d\Omega_2^{\rm L}}
=F_{\rm kin}C_0
    { \frac{d\sigma _{p\alpha}}{d\Omega _{p\alpha}}
(\theta _{p\alpha},E_{p\alpha})
}
{} \left| \bar{T}_{{\bm K}_0{\bm K}_1{\bm K}_2} \right| ^2.
\end{align}
$\theta_{p\alpha}$ is the scattering angle of the $p$-$\alpha$ binary collision
and $E_{p\alpha}$ is its scattering energy defined by
\begin{align}
E_{p\alpha}
&=\frac{\hbar ^2 {{\bm \kappa}' }^2}{2\mu_{p\alpha}},
\label{locale}
\end{align}
where $\mu_{p\alpha}$ is the reduced mass of the $p$-$\alpha$ system and ${\bm
\kappa}'$ is the asymptotic relative momentum between emitted 
{ {$p$}} and
$\alpha$-cluster in the final channel.
$F_{\rm kin}$ 
{
in Eq.~(\ref{eq:tdx})} 
is the kinematical factor defined as
\begin{align}
F_{\rm kin}&= J_{\rm L}\frac{K_1 K_2 E_1 E_2}{\hbar ^4 c^4}
\left[1+\frac{E_2}{E_{\rm{B}}}+
\frac{E_2({\bm K}_1 \cdot {\bm K}_2)}{E_{\rm{B}}K_2 ^2}\right]^{-1},
\end{align}
where $J_{\rm L}$ is the Jacobian from the center-of-mass frame to the
laboratory frame, $E_i$ and $E_{\rm{B}}$ denote the total energy of particle $i$
and  that of ${}^{6}$He, respectively. $C_0$ in Eq.~(\ref{eq:tdx}) is a
coefficient given by
\begin{align}
C_0&=\frac{E_0}{(\hbar c)^2 K_0}
\frac{\hbar ^4}{(2\pi)^3 \mu_{p\alpha}^2}{.}
\end{align}

\subsection{THSR wave function for the target and residual nuclei} \label{thsr}

We use the latest version of the THSR wave function in Ref.~\cite{zhao17} with a
di-neutron  pairing term for the descriptions of a nucleus A, which is $^{6}$He
or $^{10}$Be, as
\begin{align}
\label{eq:thsr}
  &\left| \Phi (\text{A}) \right\rangle\nonumber\\
  &\,\,\,\,\,\,=\hat{P}^J_{MK}(C_{\alpha}^{\dagger})^{(1,2)}
  \left[(1-\gamma)
   c_{n,\uparrow}^{\dagger}
   c_{n,\downarrow}^{\dagger}
   +\gamma
   c_{2n}^{\dagger}
   \right]
\left| { \bf  \rm vac} \right>.
\end{align}
Here, $\left| { \bf  \rm vac} \right>$ is the vacuum state from which the
$\alpha$ clusters and valence neutrons are created. Indices 1 and 2 in
$(C_{\alpha}^{\dagger})^{(1,2)}$ correspond to the residual nucleus ${}^{6}$He
and target nucleus ${}^{10}$Be, respectively.

$C_{\alpha}^{\dagger}$ is the creation operators of $\alpha$-clusters with the
form of
\begin{equation}
  \label{eq:alpha-creator}
  \begin{split}
  C_{\alpha}^{\dagger}&=\int d\mathbf{R}
    \exp (-\frac{R_{x}^{2}
           +R_{y}^{2}}{\beta_{\alpha,xy}^{2}}
    -\frac{R_{z}^{2}}{\beta_{\alpha,z}^{2}})\int d\mathbf{r}_{1}
      \cdots d\mathbf{r}_{4}    \\
  &\times \psi(\mathbf{r}_{1}-\mathbf{R})
      a_{\sigma_{1},\tau_{1}}^{\dagger}(\mathbf{r}_{1})
    \cdots \psi(\mathbf{r}_{4}-\mathbf{R})
      a_{\sigma_{4},\tau_{4}}^{\dagger}(\mathbf{r}_{4}),
  \end{split}
\end{equation}
where $\mathbf{R}$ is the generator coordinate of the $\alpha$-cluster and
$\psi(\mathbf{r}_{i}-\mathbf{R}) a_{\sigma_i,\tau_i}^{\dagger} (\mathbf{r}_i)$
is the single nucleon state of the $i$th nucleon in the operator form with
spin-isospin $(\sigma,\tau)$,  and with the spatial part of a Gaussian form as
$\psi(\mathbf{r}) =(\pi b^{2})^{-3/4} \exp[-r^{2}/(2b^{2})]$. The parameter $b$
is chosen to be $1.35$ fm which reproduces the size of an isolated $\alpha$
particle. $\beta_{\alpha,xy}$ and $\beta_{\alpha,z}$ are deformed parameters for
the nonlocalized motion of $\alpha$-clusters. For each nucleus, the parameters
$\beta_{\alpha,xy}$ and $\beta_{\alpha,z}$ are determined by variational
calculation.

For the valence neutrons, two types of creation operators are used with or
without the di-neutron  pairing. The creation operator $c_{n,\sigma}^{\dagger}$
for the independent configuration of valence neutrons is given by
\begin{equation}
  \label{eq:neutron-creator}
  \begin{split}
   c_{n\sigma}^{\dagger}&=\int d\mathbf{R}_{n}
    \exp \left(-\frac{R_{n,x}^{2}
           +R_{n,y}^{2}}{\beta_{n,xy}^{2}}
          -\frac{R_{n,z}^{2}}{\beta_{n,z}^{2}}\right)
    \int d\mathbf{r}_{i}    \\
  &\times (\pi b^{2})^{-3/4}
  e^{(-1)^m\phi(\mathbf{R}_{n})}
  e^{-(\mathbf{r}_{i}-\mathbf{R}_{n})^{2}/(2b^{2})}
    a_{\sigma}^{\dagger}(\mathbf{r}_{i}).
  \end{split}
\end{equation}
Here, the generator coordinate $\bm R_n$, parameters $\beta_{i,xy}$ and
$\beta_{i,z}$ for valence neutrons are  defined similarly to those in
Eq.~(\ref{eq:alpha-creator}). $a_{\sigma}^{\dagger}(\mathbf{r}_{i})$ is the
single neutron creation operator at $\mathbf{r}_i$ with spin $\sigma$. The phase
factor $\exp [(-1)^m\phi(\mathbf{R}_{n})]$ is used to describe the intrinsic
negative parity of the $\pi$-orbit state for the valence nucleons as discussed
in Refs.~\cite{lyu15,lyu16}, where $m$ is the third component of the orbital
angular momentum of the valence nucleon and $\phi(\mathbf{R}_{i})$ is the
azimuthal angle of $\mathbf{R}_i$. For the pairing configuration of valence
nucleons, the creation operator $c_{2n}^{\dagger}$ has the form 
\begin{equation}
  \label{eq:dineutron-creator}
  \begin{split}
   c_{2n}^{\dagger}&=\int d\mathbf{R}_{2n}
    \exp (-\frac{R_{2n,x}^{2}
           +R_{2n,y}^{2}}{\beta_{2n,xy}^{2}}
          -\frac{R_{2n,z}^{2}}{\beta_{2n,z}^{2}})
    \int d\mathbf{r}_{i}d\mathbf{r}_{j}    \\
  &\times (\pi b^{2})^{-3/4}
  e^{-(\mathbf{r}_{i}-\mathbf{R}_{2n})^{2}/(2b^{2})}
    a_{\uparrow}^{\dagger}(\mathbf{r}_{i})\\
  &\times (\pi b^{2})^{-3/4}
  e^{-(\mathbf{r}_{j}-\mathbf{R}_{2n})^{2}/(2b^{2})}
    a_{\downarrow}^{\dagger}(\mathbf{r}_{j}),
  \end{split}
\end{equation}
where the generator coordinate $\bm R_{2n}$, parameters $\beta_{2n,xy}$ and
$\beta_{2n,z}$ for di-neutron pair are  similarly defined to those in
Eq.~(\ref{eq:alpha-creator}). The mixture between the paring and un-pairing
configurations for the two valence neutrons are determined by the parameter
$\gamma$. Parameters $\beta_{n,xy}$, $\beta_{n,z}$, $\beta_{2n,xy}$,
$\beta_{2n,z}$ and $\gamma$ are described by the variational calculation.
Finally, we use the angular momentum projection operator $\hat{P}^J_{MK}$ to
restore the rotational symmetry of wave function \cite{schuck}.

In Eq.~(\ref{eq:thsr}), the THSR wave functions are expressed ultimately by the
creation operators of single nucleons, which consider explicitly the
single-nucleon degrees of freedom including the antisymmetrization of all the
nucleons. Because of the antisymmetrization, the formation of $\alpha$-clusters
in the inner region of the nucleus is suppressed in the present microscopic
description, even though we explicitly write the wave function of ${}^{10}$Be
with the form of $\alpha$-cluster creation operators. Hence, the so-called
spectroscopic factor $S_\alpha$ for the THSR wave function is smaller than unity
in general. The antisymmetrization effect is weak in the nuclear surface region
where the $\alpha$-knockout reaction takes place. It should be noted, however,
that the Pauli blocking effects in the inner region will affect the
$\alpha$-cluster dynamics for the entire space. Thus, only with the microscopic
description of cluster states, the amplitude of $\alpha$-cluster wave function
in the surface region can be correctly produced.

We extract the $\alpha$-cluster wave function in the surface region from the
THSR wave function of ${}^{10}$Be. This is obtained by approximating the
$\alpha$-cluster RWA using the method proposed in Ref.~\cite{kanada} which is
found to be successful in evaluating the RWA in $\alpha$-decay width
calculations. In this method, the RWA $y(a)$ at the channel radius $a$ is
approximated by $y^\textrm{app}(a)$ which is given by the overlap between the
microscopic wave function of target nucleus and a Brink-Bloch wave function as
\begin{eqnarray}
\label{eq:approx}
\left| ay(a) \right|
&\approx&
a y^\textrm{app}(a)
\nonumber \\
&\equiv&
\frac{1}{\sqrt{2}}
\left(\frac{6\times4}{10\pi b^2}\right)^{1/4}
\left|\left<
  \Phi({}^{10}\textrm{Be})|
  \Phi^{{ (0+)}}_{\textrm{BB}}({}^{6}\textrm{He},\alpha,S=a)
\right>\right|.
\end{eqnarray}
Here $\Phi({}^{10}\textrm{Be})$ is the THSR wave function of  ${}^{10}$Be, and
$\Phi_{\textrm{BB}}({}^{6}\textrm{He},\alpha,S)$ is the Brink-Bloch-type wave
function~\cite{Bri66} for the two-body system composed by the residual nucleus
${}^{6}$He and the $\alpha$-cluster with the distance parameter $S$ between
these two components, as  
\begin{equation}
    \left|\Phi^{(0+)}_{\textrm{BB}}({}^{6}\textrm{He},\alpha,S)
    \right>=
         \hat{P}^0_{00}\left|\phi(\alpha,\frac{6}{10}S\vec{e}_z)
         \Phi({}^{6} {\textrm{He}(0^+)},-\frac{4}{10}S\vec{e}_z)
         \right\rangle.        
\end{equation}
This wave function is projected onto angular momentum $J=0$ and parity $\pi=+$
eigenstates corresponding to the desired quantum numbers for the ground state of
$^{10}$Be. The THSR wave function $\Phi({}^{6}\textrm{He})$ projected into $0^+$
state is adopted in this Brink-Bloch-type wave function $\Phi_{\textrm{BB}}$ for
the description of the ${}^{6}$He component. $y^\textrm{app}(a)$ is an
approximated RWA of the $\alpha$-cluster in ${}^{10}$Be, which is regarded as
the $\alpha$-cluster wave function in the surface region.

It is already illustrated in Ref.~\cite{kanada} that the approximated RWA
$y^\textrm{app}(a)$ in Eq.~(\ref{eq:approx}) is  valid only for the outer region
of nuclei. However, benefiting from the high selectivity of the nuclear surface
region in $(p,p\alpha)$ knockout reaction \cite{yoshida16,yoshida17},  this
approximation will not affect the calculation of the $\alpha$-knockout reaction
as shown below.

\section{Results and discussion \label{secresult}}
\subsection{Numerical inputs} We study the ${}^{10}$Be($p$,$p\alpha$)${}^{6}$He
reaction at 250 MeV by taking the following kinematical conditions; the Madison
convention is adopted. The kinetic energy of particle 1 is fixed at 180 MeV and
its emission angle is set to $(\theta_1,\phi_1)=(60.9^\circ,0^\circ)$. As for
particle 2, $\phi_2$ is fixed at $180^\circ$ and $\theta_2$ is varied around
$51^\circ$ to which the kinetic energy of particle 2 changes accordingly around
62.5 MeV. For the $p$-$\alpha$ system, the scattering angle $\theta_{p \alpha}$
varies around $76^\circ$ and the scattering energy $E_{p\alpha}$ varies around
$242$~MeV. For all the scattering particles, the relativistic kinematics is
adopted.

In the microscopic calculation of the ground state wave functions of ${}^{10}$Be
and ${}^{6}$He, we follow Ref.~\cite{itagaki00} for the choice of the
nucleon-nucleon interactions; we adopt the Volkov No. 2
interaction~\cite{Volkov1965} for the central term and the G3RS
interaction~\cite{Yamaguchi1979} for the spin-orbit term. All the variational
parameters within the THSR wave function are optimized by variational
calculation for the ground state energies of ${}^{10}$Be and ${}^{6}$He.

We employ the Melbourne $g$ matrix~\cite{Amo00} as an effective NN interaction
in the folding model calculation for the optical potentials. For the
$\alpha$-${}^{6}$He potential, we adopt the target-density approximation
proposed in Ref.~\cite{Ega14}.

\subsection{Nuclear structure results} 

With the optimized THSR wave function, we obtain the energy of $-61.4$~MeV for
the ground state of ${}^{10}$Be. The extended version of the THSR wave function
described in Sec.~\ref{thsr} improves the ground state energy of $^{10}$Be by
about 1.0~MeV compared with the original one used in the previous
study~\cite{lyu16}. The corresponding root-mean-square charge radius is 2.31 fm
which agrees very well with the recent experimental value of 2.36 fm
\cite{nortershauser09}. Considering that the protons are only included in the
$\alpha$-clusters inside the ${}^{10}$Be nucleus, this good agreement shows that
the $\alpha$-cluster motion is correctly described by the THSR wave function.
{The parameters  $\beta_{\alpha,xy}$ and $\beta_{\alpha,z}$ defined in
Eq.~(\ref{eq:alpha-creator})  describe the $\alpha$-cluster motion in
${}^{10}$Be nucleus. They are optimized in the variational calculation with
values $\beta_{\alpha,xy}=0.1$~fm and $\beta_{\alpha,z}=2.6$~fm. Other
parameters are also variationally determined.}

\begin{figure}[htbp]
\centering
\includegraphics[width=0.95\columnwidth]{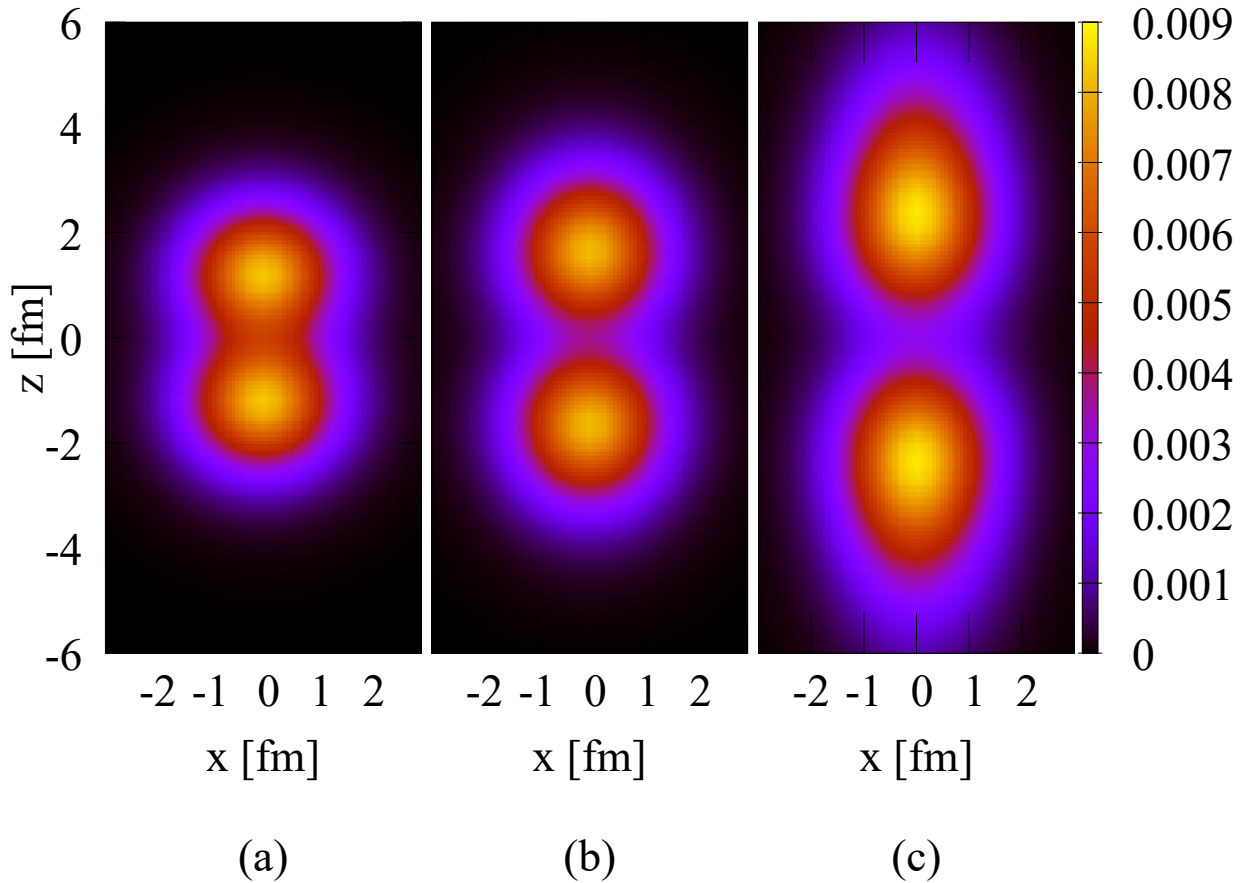}
\caption{Charge distribution of ${}^{10}$Be nucleus for the (a) artificial
shell-model like state with parameter $\beta_{\alpha,z}=1.0$~fm, (b) the
physical ground state with variational optimized parameter
$\beta_{\alpha,z}=2.6$~fm, and (c) artificial gas-like cluster state with
parameter $\beta_{\alpha,z}=6.0$~fm. (Color online) } \label{fig:dens}
\end{figure}

Fig.~\ref{fig:dens}(b) shows the intrinsic charge distribution for the cluster
structures of ${}^{10}$Be, from which a molecular like structure with two
$\alpha$-clusters located at a moderate distance is clearly observed. For
comparison, we also show the charge distribution of two artificial states of
${}^{10}$Be which are constructed by changing the parameter $\beta_{\alpha,z}$
by hand from the optimized value into two extreme values,
$\beta_{\alpha,z}=1.0$~fm and $\beta_{\alpha,z}=6.0$~fm, as shown in
Figs.~\ref{fig:dens}(a) and (c), respectively. In Fig.~\ref{fig:dens}(a), since
the $\alpha$-cluster motion is confined by the small parameter
$\beta_{\alpha,z}=1.0$~fm, a very compact distribution is observed with spatial
overlap between the two $\alpha$-clusters. In this case, the effect from
antisymmetrization is very strong and the 2$\alpha$ wave function is almost
equal to the SU(3)-shell-model limit. Hence, we call this  the shell-model like
state. On the other hand, in Fig.~\ref{fig:dens}(c), a very dilute structure is
observed for the two $\alpha$-clusters with large distance between them. Hence,
we call this  the gas-like cluster state. In Figs.~\ref{fig:dens}(a)--(c), a
continuous evolution of the nuclear structure from the shell-model limit to the
molecular-like state and then into the gas-like cluster state is observed. By
predicting the physical observables of the ${}^{10}$Be($p$,$p\alpha$)${}^{6}$He
knockout reaction, we provide experimental probes for the effective
distinguishment of these three cases of ${}^{10}$Be nucleus.

\begin{figure}[htbp]
\centering
\includegraphics[width=0.95\columnwidth]{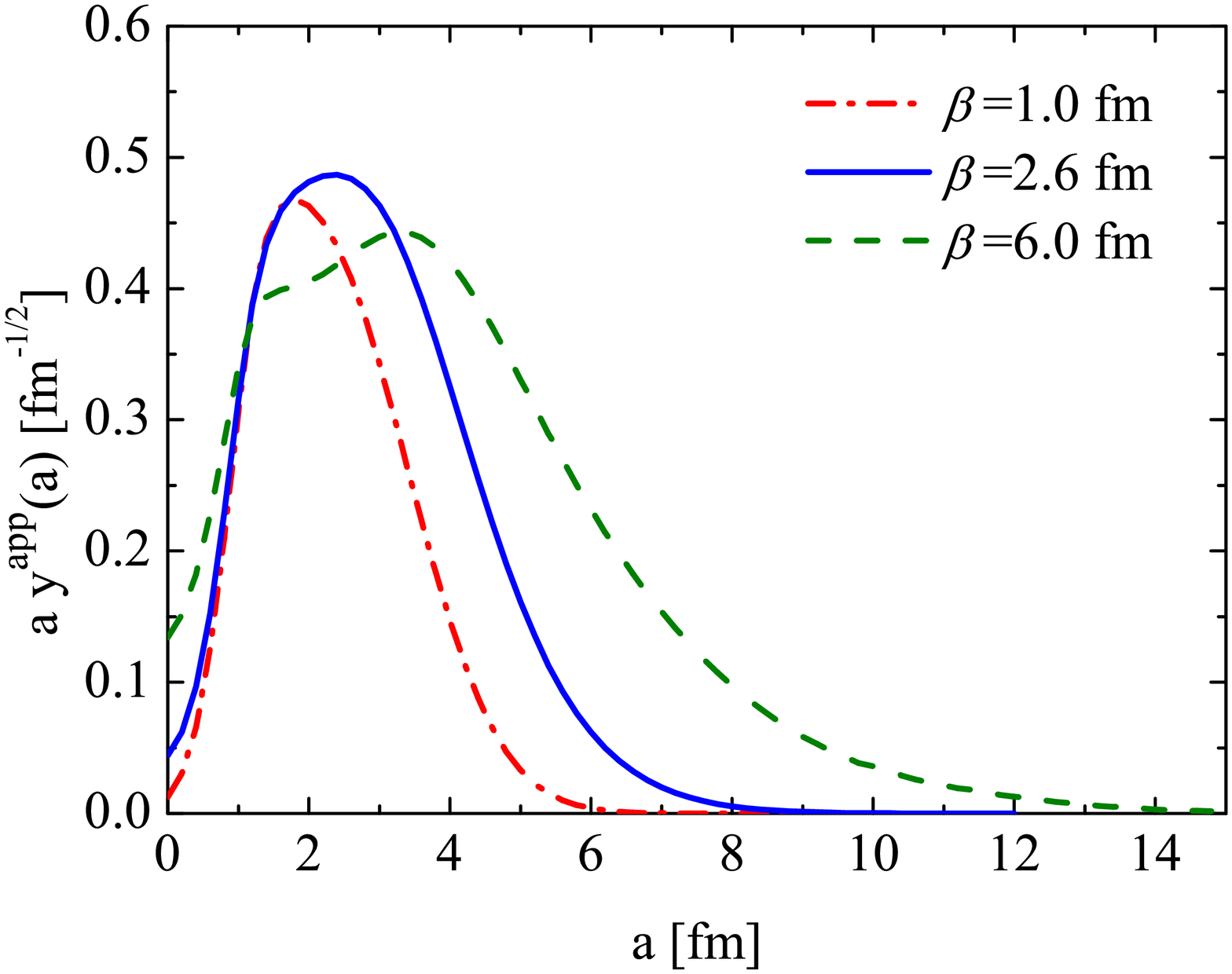}
\caption{The approximated RWA of ${}^{10}$Be nucleus for the three  values of
$\beta_{\alpha,z}$. (Color online) } \label{fig:rwa}
\end{figure}

In Fig.~\ref{fig:rwa}, we show the approximated RWA ($y^\textrm{app}(a)$)
extracted from the THSR wave function for the three cases. Clear deviations are
observed in the curves for different $\beta_{\alpha,z}$ parameters. This shows a
dramatic difference on the $\alpha$-cluster motions between the shell-model
like, molecular-like, and gas-like cluster states, as discussed above. The
present approximation of the RWA is valid in the surface and outer regions free
from the antisymmetrization, but not in the inner region, in which the RWA
should be strongly suppressed by the antisymmetrization. In the $^{10}$Be wave
function, the suppression is mainly contributed by the antisymmetrization of
nucleons between two $\alpha$-clusters. In Ref.~\cite{kanada}, it is shown that
the antisymmetrization between the two $\alpha$-clusters in ${}^{8}$Be is weak
enough  for $a\ge 3$--4~fm, which can be regarded as the physical region for two
$\alpha$-clusters. As shown later, the $\alpha$-knockout reaction can
selectively probe the clustering in the outer region, in which the RWA is safely
approximated. The $y^\textrm{app}(a)$ in the $^{10}$Be wave function in this
physical region ($a\ge 3$--4~fm) shows a clear difference of  clustering
behavior between the three states. Namely, the amplitude is remarkably enhanced
and reduced in the gas-like cluster and shell-model like states, respectively.

\subsection{Nuclear reaction results} 

\begin{figure}[t]
\centering
\includegraphics[width=0.95\columnwidth]{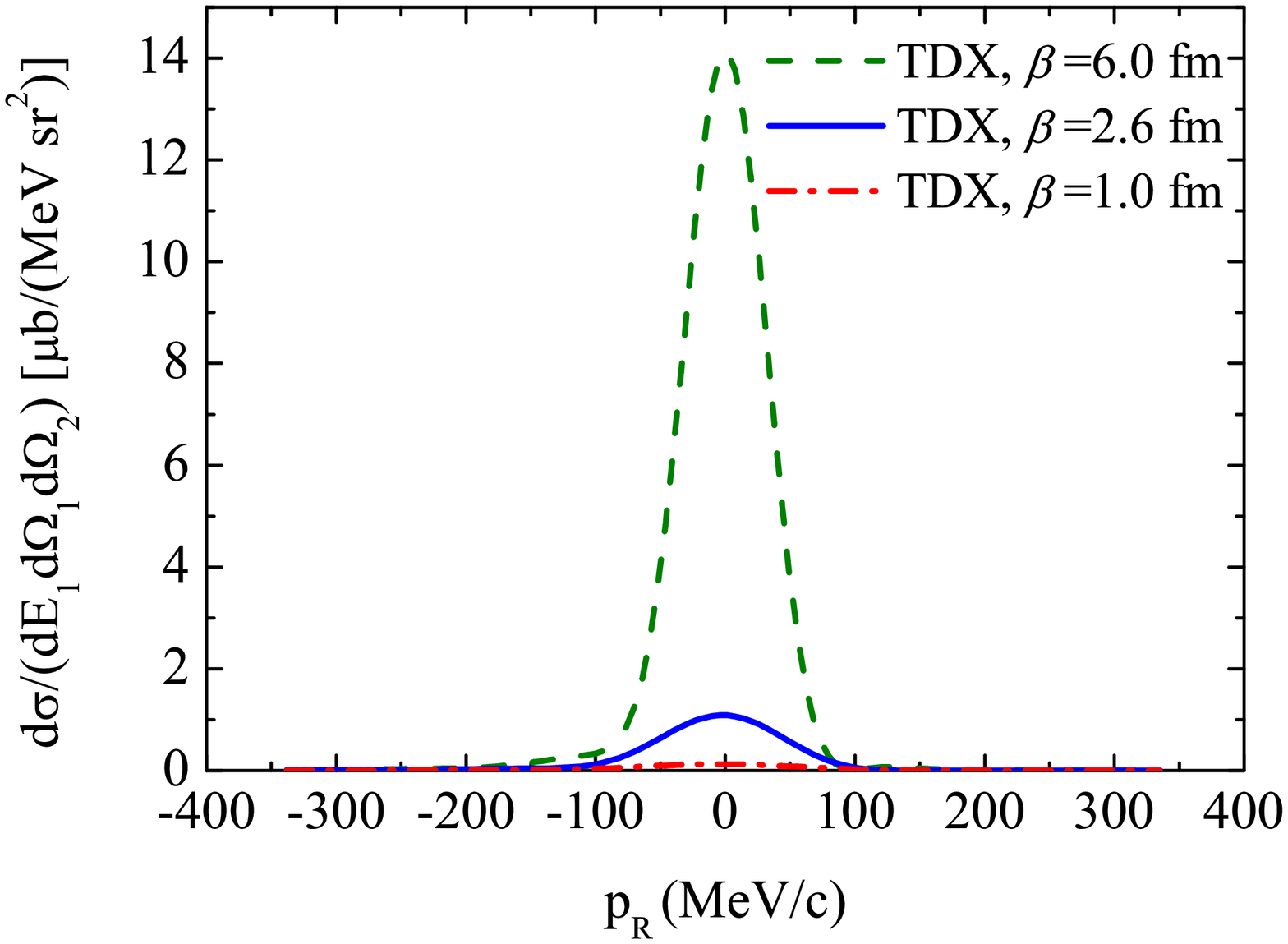}
\caption{The TDX of the ${}^{10}$Be($p$,$p\alpha$)${}^{6}$He reaction at 250
MeV. Kinetic energy of particle 1 is fixed at 180 MeV and its emission angle is
set to $(\theta_1,\phi_1)=(60.9^\circ,0^\circ)$. $\phi_2$ for particle 2 is
fixed at $180^\circ$ and $\theta_2$ is varied around $51^\circ$. $P_R$ is the
recoiled momentum. (Color online) } \label{fig:tdx}
\end{figure}

We show in Fig.~\ref{fig:tdx} the calculated TDX for the physical ground state
of ${}^{10}$Be (blue solid line) compared with the results for two artificial
states (green dashed and red dash-dotted line). We see giant ratio of about 10
for the TDXs around the peak between the shell-model like state
($\beta_{\alpha,z}=1.0$~fm) and the molecular-like state
($\beta_{\alpha,z}=2.6$~fm), and between the molecular-like state and the
gas-like cluster state ($\beta_{\alpha,z}=6.0$~fm). This shows that the TDX
observables are very sensitive to the evolution of cluster structures of the
${}^{10}$Be nucleus. By comparing the theoretical results of TDX with future
experimental values, one may easily pin down the clustering structure in the
physical ground state of  ${}^{10}$Be, which should be the molecular-like case
according to the suggestion by microscopic theories. We see very large peak of
the TDX for the gas-like cluster case and very small peak for the shell-model
like case. This is reasonable because the $\alpha$-clusters are remarkably
enhanced in the gas-like cluster case and suppressed by Pauli blocking effects
in the shell-model like case.

\begin{figure}[htbp]
\centering
\includegraphics[width=0.95\columnwidth]{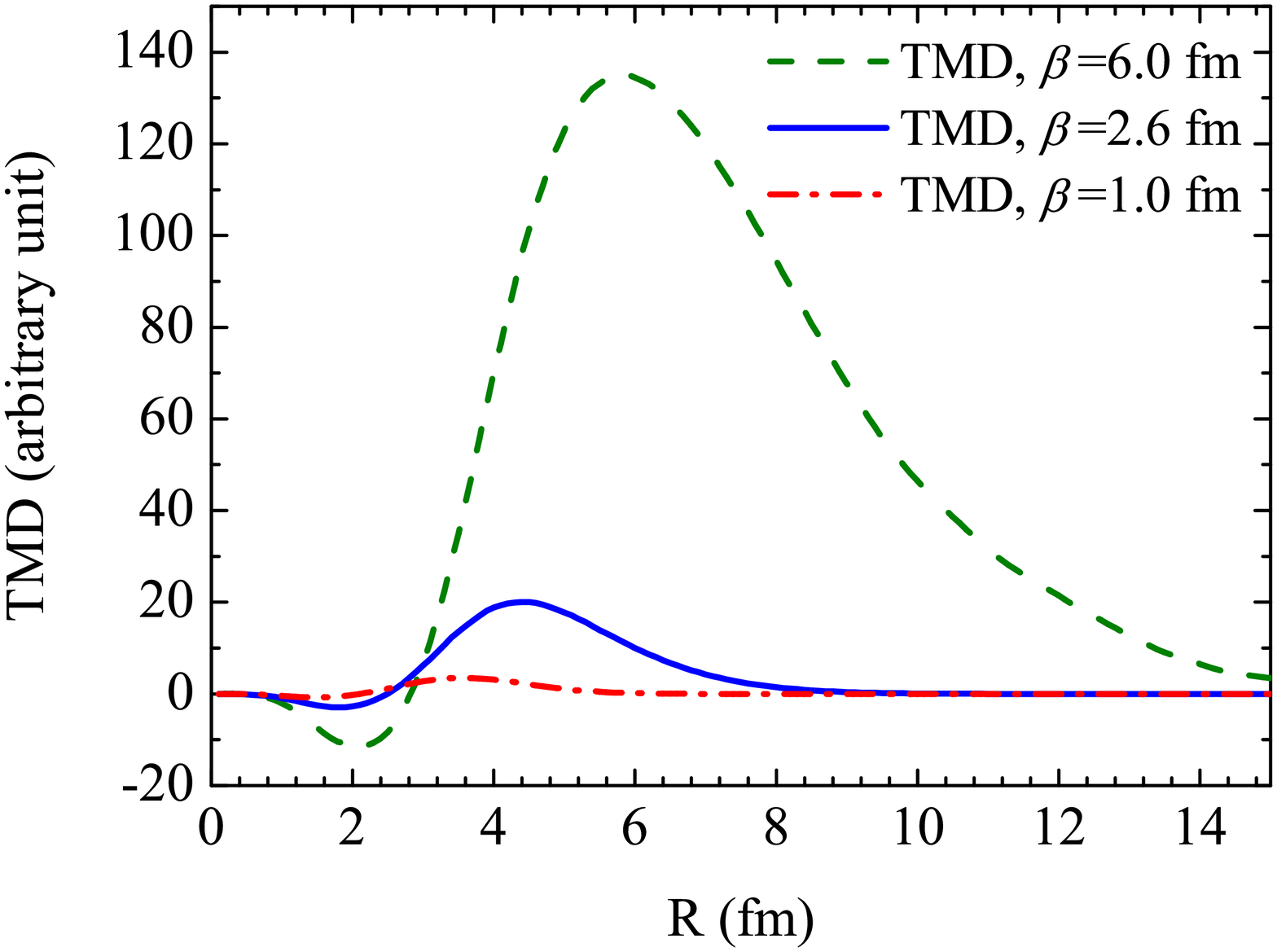}
\caption{TMD of the ${}^{10}$Be($p$,$p\alpha$)${}^{6}$He reaction at 250 MeV for
$P_R=0$. (Color online) } \label{fig:tmd}
\end{figure}

Next we discuss the peripheral property of the ${}^{10}$Be($p$,$
p\alpha$)${}^{6}$He knockout reaction by showing in Fig.~\ref{fig:tmd} the
transition matrix density (TMD) corresponding to $P_R=0$. The TMD is the
transition strength as a function of $\bm R$ as defined in Ref.~\cite{wakasa17}.
It is clearly observed that the major contributions come from the middle and
tail regions of the physical or artificial nuclei, which clearly shows the
peripheral property of the knockout reaction as discussed in
Refs.~\cite{yoshida16,yoshida17}. This is essential for  probing only the
{\lq\lq}physical'' $\alpha$-clustering in the surface and outer regions, in
which $\alpha$-clusters are free from the antisymmetrization effect, separating
from those in the inner region where the  $\alpha$-clusters are not well defined
because of the strong antisymmetrization effect. As mentioned above, the
approximated $\alpha$-cluster RWA is reliable only in the surface and outer
regions. Nevertheless, we may conclude from Fig.~\ref{fig:tmd} that the TMD has
significant distribution only in the outer region where the approximated RWA
adopted is reliable. Therefore, it will be expected that the features of the TDX
for three states in Fig.~\ref{fig:tdx} persist.

In previous studies of clustering physics, it is difficult to obtain direct
experimental evidence of $\alpha$-clustering. With our microscopic
$\alpha$-knockout reaction framework, we are able to detect directly the
$\alpha$-clustering in the surface region of nuclei. Furthermore,
distinguishment between the compact molecular-like and gas-like cluster states
is also possible in this new framework.

\section{Summary} \label{secsum} 

We have proposed the first  microscopic framework for the study of ($p,p\alpha$)
knockout reaction by integrating a microscopic clustering model into the DWIA
framework. With this new framework, we investigated the $\alpha$-knockout
reaction ${}^{10}$Be($p$,$p\alpha$)${}^{6}$He at 250 MeV. The target nucleus
${}^{10}$Be and the residual nucleus ${}^{6}$He in this reaction are described
microscopically by the THSR wave function. An approximated $\alpha$-cluster RWA
$y^\textrm{app}(a)$ is extracted from the THSR wave function of target nucleus
${}^{10}$Be following Ref.~\cite{kanada}, and implemented in the reaction
calculation. By predicting the TDX for the ${}^{10}$Be($p$,$p\alpha$)${}^{6}$He
reaction, we have provided possibility for the direct manifestation of the
$\alpha$-clustering in ${}^{10}$Be. We also compared the structures and reaction
observables for the physical and two artificial states of the target nucleus
${}^{10}$Be, namely the molecular-like, shell-model limit, and gas-like cluster
states. The $\alpha$-cluster amplitudes for these three states are very
different from each other. In consequence of this, we observed giant ratio
between  their reaction observable TDXs, which shows the strong dependence of
the TDX on the $\alpha$-clustering structure in ${}^{10}$Be. Another important
finding is the peripheral property of the knockout reaction, which guarantees
high selectivity for probing the $\alpha$-cluster in the surface region and
allows one to use $y^\textrm{app}(a)$.  Using our framework, we may directly
relate the microscopic description of $\alpha$-clustering structure to the
reaction observables in the ($p,p\alpha$) knockout reaction, and provide
sensitive manifestation of $\alpha$-clustering in the ${}^{10}$Be nucleus. In
future, it is appealing and hopeful to extend this microscopic framework for
systematic studies of  $\alpha$-clustering states.

\begin{acknowledgments}
The authors thank K.~Minomo, Y.~Chazono, and N.~Itagaki for valuable
discussions. The computation was carried out with the computer facilities at the
Research Center for Nuclear Physics, Osaka University. This work was supported
in part by Grants-in-Aid of the Japan Society for the Promotion of Science
(Grants No. JP16K05352 and No. JP15J01392).
\end{acknowledgments}


\end{document}